\documentclass[aps,pra,amsmath,amssymb,floatfix,twocolumn,amsmath,superscriptaddress,twocolumn,nofootinbib,tighten,letterpaper]{revtex4-2}
\usepackage[colorlinks,linkcolor=blue,citecolor=blue,urlcolor=blue]{hyperref}
\usepackage{multirow}
\usepackage{subfigure}
\usepackage{color}
\usepackage{mathrsfs}
\usepackage{hyperref}
\usepackage[normalem]{ulem}
\usepackage{bm}

\usepackage{amssymb}
\usepackage{amsmath}
\renewcommand\vec[1]{\ensuremath\boldsymbol{#1}}

\usepackage{tabularray}

\usepackage{array} 
\newcolumntype{P}[1]{>{\centering\arraybackslash}p{#1}}

\definecolor{RowColor}{rgb}{0.88,1,0.9}

\usepackage{amsfonts, relsize, color, mathtools, physics}
\usepackage{graphics}
\usepackage{graphicx}
\usepackage{subfigure}
\usepackage{color}
\usepackage{comment}

\begin{document}

\title{Charge density waves and stripes in quarter metals of graphene heterostructures}

\author{Sk Asrap Murshed}
\affiliation{Department of Physics, Lehigh University, Bethlehem, Pennsylvania, 18015, USA}

\author{Bitan Roy}~\thanks{Contact author: bitan.roy@lehigh.edu}
\affiliation{Department of Physics, Lehigh University, Bethlehem, Pennsylvania, 18015, USA}

\date{\today}

\begin{abstract}
 Motivated by recent experiments, here we identify valley-coherent charge density wave (VC-CDW) in the nondegenerate quarter metal for the entire family of chirally-stacked $n$ layer graphene, encompassing rhombohedral multilayer, Bernal bilayer, and monolayer cousins. Besides the hallmark broken translational symmetry, yielding a modulated charge density over an enlarged unit-cell with a characteristic $2{\bf K}$ periodicity, where $\pm {\bf K}$ are the valley momenta, this phase lacks the three-fold ($C_3$) rotational symmetry but only for $n=1$ and even integer $n$. The VC-CDW then represents a stripe order, as observed in hexalayer graphene [\href{https://arxiv.org/abs/2504.05129}{arXiv:2504.05129}], but preserves the $C_3$ symmetry for other odd integer $n$ as observed in trilayer graphene [\href{https://www.nature.com/articles/s41567-024-02560-7}{Nat.\ Phys.\ {\bf 20}, 1413 (2024)} and \href{https://arxiv.org/abs/2411.11163}{arXiv:2411.11163}]. From a \emph{universal} Clifford algebraic argument, we show that the VC-CDW and an anomalous Hall order can lift the residual valley degeneracy of an antiferromagnetically ordered spin-polarized half metal, when these systems are subject to perpendicular displacement fields. Only the anomalous Hall order displays a hysteresis in off-diagonal resistivity, as observed in all the systems with $2 \leq n \leq 6$. We showcase a confluence of VC-CDW and anomalous Hall orders within the quarter metal, \emph{generically} displaying a regime of coexistence, otherwise separating the pure phases.
\end{abstract}

\maketitle

\section{Introduction}

\begin{figure}[t!]
    \includegraphics[width=0.95\linewidth]{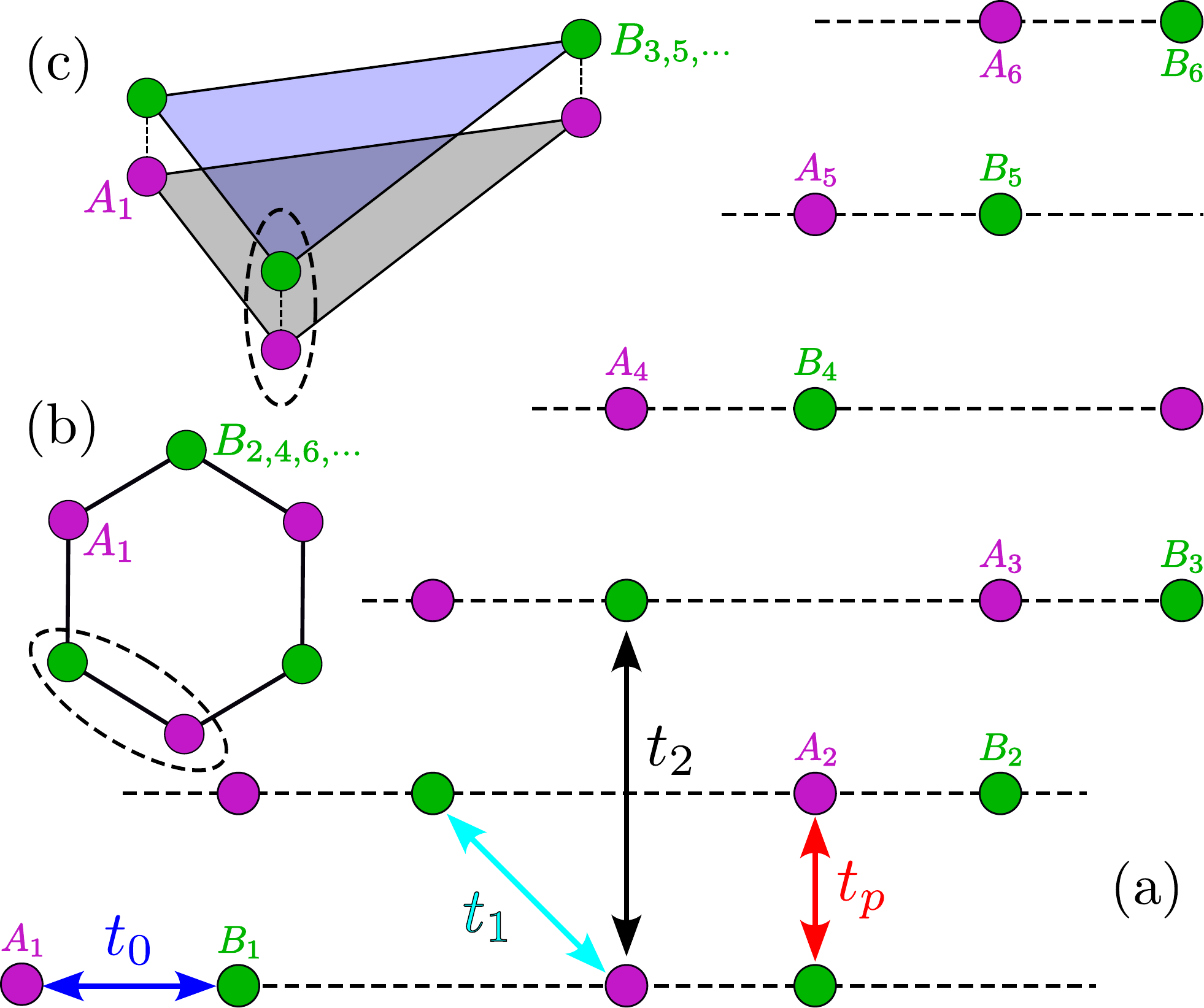}
    \caption{(a) Lattice structure of chirally-stacked $n$ layer graphene, where $A_j$ (magenta) and $B_j$ (green) correspond to the sites belonging to the $A$ and $B$ sublattice of the honeycomb lattice on the $j$th layer. In a system with $n$ layers, the low-energy two-band model near each valley and for each spin projection is constituted by the Wannier states localized on $A_1$ and $B_n$ sites. Hopping processes are denoted by double-headed arrows. The emergent lattice, constituted by low-energy sites, form (b) a honeycomb lattice when $n=2,4,6, \cdots$ and (c) a prism-like lattice for $n=3,5,\cdots$, where the two-site unit cells are shown by dashed closed loops.   
    }~\label{fig:lattice}
\end{figure}

A periodic modulation of charge, realized at the cost of the underlying lattice translational symmetry gives rise of a novel quantum state of matter, known as the charge density wave (CDW). Such a phase can result from electron-phonon interactions in weakly correlated systems or from electronic repulsions in correlated materials. Furthermore, when such a density modulation breaks crystallographic discrete rotational symmetry, an analog of the smectic order is realized in electronic liquids, also known as the stripe phase~\cite{CDWSTRIPE:1, CDWSTRIPE:2, CDWSTRIPE:3, CDWSTRIPE:4, CDWSTRIPE:5, CDWSTRIPE:6, CDWSTRIPE:7, CDWSTRIPE:8}. In this work, we identify the footprints of such CDW and stripe orders in the so called quarter metal phase, devoid of the spin and valley degeneracies, in chirally-stacked $n$ layer graphene, a family that encompasses monolayer ($n=1$), Bernal bilayer ($n=2$), and rhombohedral multilayer ($n \geq 3$) graphene heterostructures (Fig.~\ref{fig:lattice}). We systematically classify symmetry-breaking orders using the Clifford algebraic structure of the graphene Hamiltonian and show how mutually commuting orders sequentially lift degeneracies, thereby identifying valley-coherent charge density wave (VC-CDW) and anomalous Hall order (AHO) as candidates for the experimentally observed quarter metal in multilayer graphene.

The rest of the paper is organized as follows. In Sec.~\ref{sec:summary}, we summarize the key findings of this work. Experimental status of the field is briefly reviewed in Section~\ref{sec:experiments}. A minimal model Hamiltonian for free fermions in chirally stacked $n$ layer graphene heterostructures in the presence of externally-induced layer or sublattice polarization is shown in Section~\ref{sec:freefermions}. Section~\ref{sec:fractionalmetals} is devoted to the discussion on systematic spin and valley degeneracy lifting, and sequential formation of half metal and quarter metal. Discussions on our findings and possible future directions are presented in Sec.~\ref{sec:discussions}.

\section{Summary of key results}~\label{sec:summary}

We show that the CDW always breaks the translational symmetry, leading to a charge modulation that is commensurate in an enlarged unit-cell (Fig.~\ref{fig:orders}) with a characteristic $2{\bf K}$ periodicity, where $\pm {\bf K}$ are the valley momenta, thus representing a valley-coherent order. Additionally it breaks the three-fold rotational ($C_3$) symmetry, but only for $n=1,2,4,6,\cdots$ yielding a stripe phase therein as observed in hexalayer graphene~\cite{ref:RHLG:1}. By contrast, the $C_3$ symmetry remains preserved in such a phase when $n=3,5, \cdots$ as observed via the scanning tunneling microscopy in trilayer graphene~\cite{ref:RTLG:5, ref:RTLG:6}. Our proposed \emph{universal} mechanism for the cascade of degeneracy lifting due to the simultaneous presence of competing orders, ultimately leading to the identification of such a VC-CDW order in the quarter metal, rests on the following robust and general Clifford algebraic argument.

If the electronic bands possess a $2^N$-fold degeneracy, then its $2^M$-fold degeneracy can be lifted in the simultaneous presence of $M+1$ number of orders for which the corresponding matrix operators mutually \emph{commute}, where $M \leq N$ and $N=2$ due to the valley and spin degrees of freedom for any $n$. Thus, in a half metal (quarter metal), requisite $2$-fold ($4$-fold) degeneracy lifting is accomplished in the simultaneous presence of two (three) mutually commuting orders. This is the \emph{first} Clifford algebraic criterion that selects unique sets of orders, capable of lifting the spin and valley degeneracies from electronic bands, which we exemplify with specific orders later in the work. Furthermore, once a set of orders is picked for either the spin or the valley degeneracy lifting following the first Clifford algebraic criterion, a \emph{second} one allows us to organize them according to their propensity toward the nucleation on an energetic ground in the following way. Among such orders, the ones that fully \emph{anticommute} with the momentum-dependent kinetic terms, excluding the perpendicular displacement ($D$) field-induced layer or sublattice polarization and chemical potential, are energetically most favored and are named \emph{mass}. Orders that only partially commute or anticommute with the kinetic terms are the next best candidates energetically, but generally less favorable and therefore less likely to occur. By contrast, the ones that fully commute with the kinetic term are energetically most inferior. We exemplify such a competition in more detail with specific examples later in the manuscript. Invoking these two Clifford algebraic criteria, we arrive at the following conclusions. Specifically, we recognize that in the presence of $D$ field-induced layer polarization, an antiferromagnetic order causes spin degeneracy lifting, yielding a valley-degenerate half metal, while an additional formation of the VC-CDW or an AHO gives rise to the nondegenerate quarter metal, as typically observed in experiments.

Finally, a third Clifford-algebra criterion shows that when several mutually anticommuting orders lift the same band degeneracy (spin or valley, for example), their coexistence is energetically favored because then their amplitudes enter the quasiparticle spectrum as a sum of their squares. We illustrate this generic mechanism through the competition between the AHO and VC-CDW orders in the quarter metal phase (Fig.~\ref{fig:meanfield}).

It should be noted that all the ordered states are described by \emph{product} matrices operative over the independent or decoupled layer or sublattice, valley, and spin degrees of freedom, which thus either mutually commute or anticommute. In the literature~\cite{symmbreak:1, symmbreak:2}, competing orders are somewhat exclusively discussed for mutually anticommuting matrices. As their amplitudes enter the quasiparticle dispersion as the sum of their squares, simultaneous presence of such mutually anticommuting competing orders does not cause any degeneracy lifting of electronic bands, which can only be accomplished in the simultaneous presence of mutually commuting orders. Thus, our proposed Clifford algebraic criterion for the formation of half metal and quarter metal is \emph{definitive} in the sense that (a) the first Clifford algebraic criterion selects unique sets of orders for the spin and valley degeneracy lifting, (b) the second Clifford algebraic criterion suggests that among such preselected orders the ones corresponding to mass are most prominent, while the other candidates typically display a suppressed propensity toward nucleation and thus appear less commonly in the global phase diagram, and (c) in the presence of mutually anticommuting orders, the third Clifford algebraic criterion favors a regime of their coexistence in between the pure phases, within a given fractional metal.

\section{Experimental observations}~\label{sec:experiments}

Graphene-based electronic liquids are endowed with valley and spin degeneracies, raising the possibility of realizing a plethora of ordered phases that break some discrete and/or continuous symmetries of the normal state~\cite{symmbreak:1, symmbreak:2, symmbreak:3, symmbreak:4, symmbreak:5}. Such ordering tendencies receive an enhanced propensity with increasing number of layers ($n$) in chirally-stacked heterostructures, as the density of states therein scales as $|E|^{2/n-1}$. A wave of recent experiments showed clear signatures of a cascade of spontaneous symmetry breaking in these systems with $2 \leq n \leq 6$, when they are subject to $D$ fields and gradually doped away from the charge-neutrality point (CNP)~\cite{ref:RHLG:1, ref:RTLG:5, ref:RTLG:6, ref:BBLG:1, ref:BBLG:2, ref:BBLG:3, ref:BBLG:4, ref:RTLG:2, ref:RTLG:3, ref:RTLG:4, ref:TTLG:1, ref:TTLG:2, ref:TTLG:3, ref:TTLG:4, ref:RPLG:1, ref:RPLG:2, ref:RPLG:3, ref:RHLG:5, ref:RHLG:6, ref:RHLG:7, ref:RHLG:8}. By contrast, electronic liquid in monolayer graphene continues to feature nodal Dirac fermions without any ordering due to the linearly vanishing density of states therein~\cite{ref:MLG:1, ref:MLG:2, ref:MLG:3}. However, recent experiments have reported relativistic Mott transitions in designer graphene or Dirac materials~\cite{ref:MLG:4, ref:MLG:5}.

Experimental findings allow us to construct a global phase diagram for such systems with $2 \leq n \leq 6$. At relatively high doping, the system is described by a metallic phase, possessing four-fold valley and spin degeneracy. At moderate doping the system typically loses the spin degeneracy, yielding a metallic state with only two-fold valley degeneracy, named half metal. Then at low doping the residual valley degeneracy gets lifted, yielding a nondegenerate quarter metal. Nonetheless, a valley-polarized half metal is also conceivable, as observed recently in pentalayer ($n=5$) rhombohedral graphene~\cite{ref:RPLG:3}. Systematic degeneracy lifting from electronic bands is clearly recognized from quantum oscillation frequencies. Such a systematic degeneracy lifting yielding fractional metals results from spontaneous breakdown of symmetries, caused by the nucleation of various ordered states in the system. Furthermore, superconductivity has been observed in the global phase diagram, primarily near the metal and the half metal. These experimental observations caused a surge of theoretical works geared toward unfolding the nature and microscopic origins of various ordered phases and the pairing symmetries~\cite{ref:RTLGnew:1, ref:RTLGnew:2, ref:RTLGnew:3, ref:RTLGnew:4, ref:RTLGnew:5, ref:RTLGnew:6, ref:RTLGnew:7, ref:RTLGnew:8, ref:RTLGnew:10, ref:RTLGnew:11, ref:RTLGnew:12, ref:RTLGnew:13, ref:RTLGnew:14, ref:RTLGnew:16, ref:RTLGnew:18, ref:RTLGnew:19, ref:BBLGRTLGnew:3, ref:BBLGRTLGnew:7, ref:BBLGnew:1, ref:BBLGnew:3, ref:BBLGnew:4, ref:BBLGnew:5, ref:BBLGnew:6, ref:BBLGnew:8, ref:BBLGnew:9, ref:BBLGRTLGTTLGnew:3, ref:BBLGRTLGTBLGnew:1, ref:MLGBBLGRTLGnew:1, ref:RTLGRHLGnew:1, ref:RNLGnew:1}. After a theoretical proposal~\cite{ref:BBLGRTLGnew:7}, more recently superconductivity in the close proximity to the quarter metal has also been observed~\cite{ref:RHLG:1, ref:TTLG:1, ref:TTLG:4}. Due to the spin- and valley-polarized nature of the quasiparticles in the quarter metal, the resulting paired state must correspond to an odd-parity chiral pair density wave, triggering a wave of new theoretical interest~\cite{ref:TTLGnew:1, ref:TTLGnew:3, ref:TTLGnew:5, ref:TTLGnew:6, ref:TTLGnew:7, ref:TTLGnew:8, ref:TTLGnew:9}.

\section{Free fermion model}~\label{sec:freefermions}

We consider a minimal tight-binding description for free-fermions in chirally-stacked $n$ layer graphene (Fig.~\ref{fig:lattice}). We take into account hopping amplitudes between the sites from two sublattices living on the same layer ($t_0$), yielding massless Dirac fermions and on the adjacent layers ($t_1$), yielding a momentum-dependent trigonal warping. We also account for two vertical hopping amplitudes between the sites from different sublattices, otherwise living on the adjacent layers ($t_p$), responsible for the band splitting and on the second-adjacent layer ($t_2$), yielding a momentum-independent trigonal warping~\cite{ref:MLGth:1, ref:BBLGth:1, ref:RTLGth:1, ref:RPLGth:1, ref:GrphnRMP:1}. We neglect hopping between the sites from the same sublattice, yielding a momentum-dependent particle-hole asymmetry. It can be absorbed in the redefinition of the chemical potential ($\mu$), measured from the CNP. The $D$ field enters the Hamiltonian in terms of an on-site potential on the $m$th layer of an $n$-layer system $V_{nm}=V[1/2-(m-1)/(n-1)]$ with $m \leq n$, where $V=Dd(n-1)$ and $d$ is the inter-layer distance.

The Bloch Hamiltonian in all the systems at a lattice momentum $\vec{q}$ can then be cast as a block matrix
\begin{equation}
H_{n \: {\rm layer}} (\vec{q}) = \left( \begin{array}{cc}
H_{LL} & H_{LH} \\
H_{HL} & H_{HH}
\end{array}
\right) (\vec{q}),
\end{equation}
where $H_{LL} (\vec{q})$ is a $2 \times 2$ matrix, operative over the low-energy sites ($A_1$ and $B_n$) and $H_{HH} (\vec{q})$ is a $(2 n-2) \times (2n-2)$ matrix acting on the high-energy sites ($A_{2,\cdots,n}$ and $B_{1, \cdots,n-1}$) on which the split-off bands live predominantly. Here, $A_m$ ($B_m$) corresponds to the sites belonging to the $A$ ($B$) sublattice on the $m$th layer. See Fig.~\ref{fig:lattice}. The coupling between these two sets of sites is captured by $H_{LH}(\vec{q})$ and $H_{HL}(\vec{q}) \equiv H^\dagger_{LH}(\vec{q})$. To arrive at the effective low-energy description, we integrate out the high-energy sites, leading to the renormalized Hamiltonian $H^{\rm renor}_{n \: {\rm layer}} (\vec{q}) = H_{LL}(\vec{q}) - H_{LH} H^{-1}_{HH} H_{HL} (\vec{q})$~\cite{symmbreak:2, BandProj:1, BandProj:2}. The continuum model is derived by expanding each component of $H^{\rm renor}_{n \; {\rm layer}} (\vec{q})$ to the leading order in a small momentum $\vec{k}$ around each valley with $\vec{q}= \pm {\bf K}+\vec{k}$. Inclusion of the spin degrees of freedom leads to a mere doubling due to a negligibly small spin-orbit coupling for light carbon atoms. We then arrive at the final form of the low-energy Hamiltonian $H^{\rm renor, low}_{n \; {\rm layer}} (\vec{k})=\sigma_0 \otimes \left[ H_{n}(\vec{k}) -\mu \right]$, where $\{ \sigma_\nu \}$ is a set of Pauli matrices operating on the spin index with $\nu=0, \cdots, 3$, $\otimes$ stands for the Kronecker product, and $H_{n}(\vec{k})=\Gamma_{kj} \; d^{n}_{kj} (\vec{k}) + \Gamma_{03} d^{n}_{03} (\vec{k})$ with $k=0,3$ and $j=1,2$. A summation over the repeated indices is assumed. Lengthy expressions for $d^{n}_{\nu \rho} (\vec{k}) \equiv d^{n}_{\nu \rho}$ (for brevity) are not instructive, but displayed in the Supplemental Material (SM)~\cite{SM}. Four-component Hermitian matrices are $\Gamma_{\nu \rho}= \tau_\nu \otimes \eta_\rho$, where the set of Pauli matrices $\{ \tau_\nu \}$ ($\{ \eta_\nu \}$) operates on the valley (sublattice or layer) index. Due to the valley inversion symmetry, generated by $\sigma_0 \otimes \Gamma_{10}$ under which $\vec{k} \to (-k_x,k_y)$~\cite{ref:MLG:2}, for any momenta either $d^n_{01,32}(\vec{k})$ or $d^n_{31,02}(\vec{k})$ are finite.

\begin{figure}[t!]
    \includegraphics[width=0.95\linewidth]{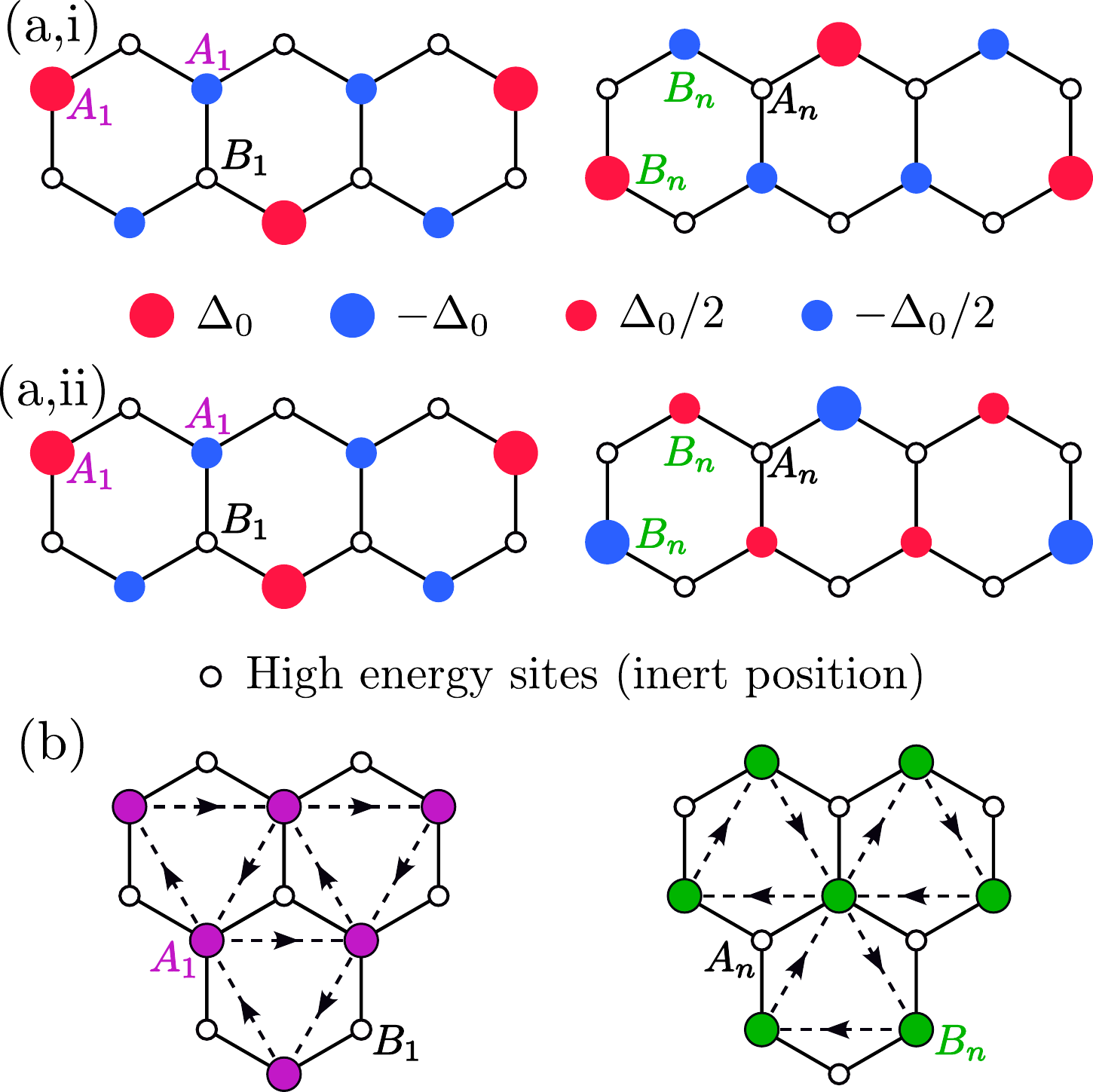}
    \caption{(a) Lattice realization of the valley-coherent charge density wave of real amplitude $\Delta_0$ with (i) in-phase and (ii) out-of-phase modulations between the sites from the $A_1$ and $B_n$ sublattices (Fig.~\ref{fig:lattice}). (b) Intra-sublattice/layer circulating currents in the direction of arrows, yielding the anomalous Hall order. The high-energy inert $B_1$ ($A_n$) sites in the bottom (top) layer are represented by open black circles. For $n>2$, all the sites in intermediate layers are inert (not shown here).       
    }~\label{fig:orders}
\end{figure}

\section{Metal and fractional metals}~\label{sec:fractionalmetals}

While $d^{n}_{01,02,31,32}$ account for all the hopping processes, $d^{n}_{03}$ stems from the $D$ field, causing a layer polarization of electronic density and its $k$-dependence gives rise to a sagging of electronic bands, yielding annular (simply connected) Fermi rings at low (moderate) doping. As the $d^{n}_{03}$ term anticommutes with the rest of the matrices in $H_{n}(\vec{k})$, the resulting layer polarization stands as an externally-induced mass order for gapless chiral fermions, producing an isotropic gap near the CNP. At this stage, the electronic bands are four-fold degenerate, with the energy spectrum $\pm [(d^n_{01/31})^2 + (d^n_{32/02})^2 + (d^n_{03})^2]^{1/2}-\mu$, where $\pm$ corresponds to conduction and valence bands, respectively. Hence, for any $\mu > {\rm min}.(d^n_{03})$ the system describes a four-fold degenerate metal. Next we search for possible channels of symmetry breaking, leading to the formation of fractional metals.

\subsection{Spin-polarized and valley-degenerate half metal}

Following the first Clifford algebraic principle of degeneracy lifting, outlined earlier, we proceed to search for an order for which the matrix operator \emph{commutes} with $\sigma_0 \otimes \Gamma_{03}$ (accompanying $d^n_{03}$), such that its simultaneous presence with the $D$ field induced layer-polarized mass order lifts the spin degeneracy from electronic bands. We note that among various competing orders, nucleation of the ones for which the matrix operator fully \emph{anticommutes} with the Hamiltonian for gapless chiral fermions yields a maximal gain in the condensation energy as then all the states below the Fermi energy get pushed down uniformly (second Clifford algebraic criterion). Such orders are named \emph{mass}. With the inclusion of the on-site Hubbard repulsion ($U$), the dominant component among all the finite-range interactions, these systems acquire a propensity toward the formation of an antiferromagnetic mass order. Such an order is represented by $\vec{\sigma} \otimes \Gamma_{03}$, featuring a staggered pattern of electronic spin between two layers/sublattices~\cite{symmbreak:5, ref:RTLGnew:10, ref:BBLGnew:3}. As $[\sigma_0 \otimes \Gamma_{03}, \vec{\sigma} \otimes \Gamma_{03}]=0$, the condensation of the antiferromagnetic order (with an amplitude $|{\bf \Delta}_0|=\Delta_0$) in the presence of $D$ field induced layer polarization lifts the spin degeneracy of electronic bands, leaving their valley degeneracy unaffected (as both orders appear with $\tau_0$). This outcome can be appreciated from the eigenspectrum of the associated effective single-particle Hamiltonian $H^{\rm renor, low}_{n \: {\rm layer}} (\vec{k}) +{\bf \Delta}_0 \cdot \vec{\sigma} \otimes \Gamma_{03}$, given by $\pm [(d^n_{01/31})^2 + (d^n_{32/02})^2 + \{ d^n_{03} \pm \Delta_0 \}^2 ]^{1/2}-\mu$. The system features a spin-polarized, valley-degenerate half metal when $d^n_{03}(0) - \Delta_0 < \mu < d^n_{03}(0) + \Delta_0$.

It should be noted that the first Clifford algebraic argument also allows a \emph{ferromagnetic} order to lift the spin degeneracy from electronic bands without affecting their valley degeneracy, for which the matrix operator is $\vec{\sigma} \otimes \Gamma_{00}$, appearing with the trivial identity matrix $\tau_0$ in the valley space and satisfying the requisite commutation relation $[\sigma_0 \otimes \Gamma_{03}, \vec{\sigma} \otimes \Gamma_{00}]=0$. However, $\vec{\sigma} \otimes \Gamma_{00}$ \emph{commutes} with all the matrices appearing in the momentum-dependent kinetic terms ($\Gamma_{kj}$ with $k=0,3$ and $j=1,2$). Consequently, a ferromagnetic order is energetically far less favorable than the antiferromagnetic mass order, according to the second Clifford algebraic criterion. This may explain its absence in the global phase diagrams of  chirally-stacked $n$ layer graphene heterostructures, consistent with the lack of magnetic hysteresis in their half metallic phases.

\subsection{Fully polarized nondegenerate quarter metal}

We follow the road map of the first Clifford algebraic argument for the systematic degeneracy lifting. Next we search for orders that would lift the residual valley-degeneracy of the half metal, leading to the formation of a quarter metal at lower chemical doping, devoid of any degeneracy. Naturally, we must search for the orders whose matrix operator commutes with the ones for layer polarization ($\sigma_0 \otimes \Gamma_{03}$) and antiferromagnet ($\vec{\sigma} \otimes \Gamma_{03}$) [or ferromagnet ($\vec{\sigma} \otimes \Gamma_{00}$)]. If we restrict such a search within the set of mass orders (second Clifford algebraic criterion), then there exists a \emph{unique} candidate $\sigma_0 \otimes \Gamma_{33}$ which corresponds to the AHO, resulting from intra-sublattice/layer circulating current, originally proposed by Haldane~\cite{haldane}, see Fig.~\ref{fig:orders}(b). The intra-layer next-nearest-neighbor Coulomb repulsion ($V_2$)~\cite{symmbreak:5, ref:RTLGnew:10, ref:BBLGnew:3}, the next dominant component among all finite range interactions in chirally-stacked layered graphene systems, among spinless fermions (due to the formation of precursor spin-polarized half metal) is a natural microscopic source for such an order that causes valley polarization (appearing with $\tau_3$). As such an order yields a finite orbital magnetic moment, its nucleation can be identified from a finite anomalous Hall conductivity ($\sigma_{xy}$) and a hysteresis in the off-diagonal resistivity ($R_{xy}$), which have been observed in all systems with $2 \leq n \leq 6$ so far. The valley-degeneracy lifting due to the AHO (with an amplitude $\Delta_1$) can be appreciated from the eigenspectrum of the effective single-particle Hamiltonian $H_n(\vec{k})-\Delta_0 \Gamma_{03} -\Delta_1 \Gamma_{33}-\mu$ for spinless fermions, describing the spin-polarized half metal, in the simultaneous presence of $D$ field induced layer-polarization, antiferromagnet, and AHO, given by $\pm [(d^n_{01/31})^2 + (d^n_{32/02})^2 + \{ |d^n_{03} - \Delta_0| \pm \Delta_1 \}^2 ]^{1/2}-\mu$. Therefore, when $|d^n_{03}(0) - \Delta_0| -\Delta_1 < \mu < |d^n_{03}(0) - \Delta_0| +\Delta_1$ (assuming $\Delta_0>\Delta_1$ which is justified as $U>V_2$), the system features a fully polarized quarter metal.

\begin{figure}[t!]
    \includegraphics[width=0.95\linewidth]{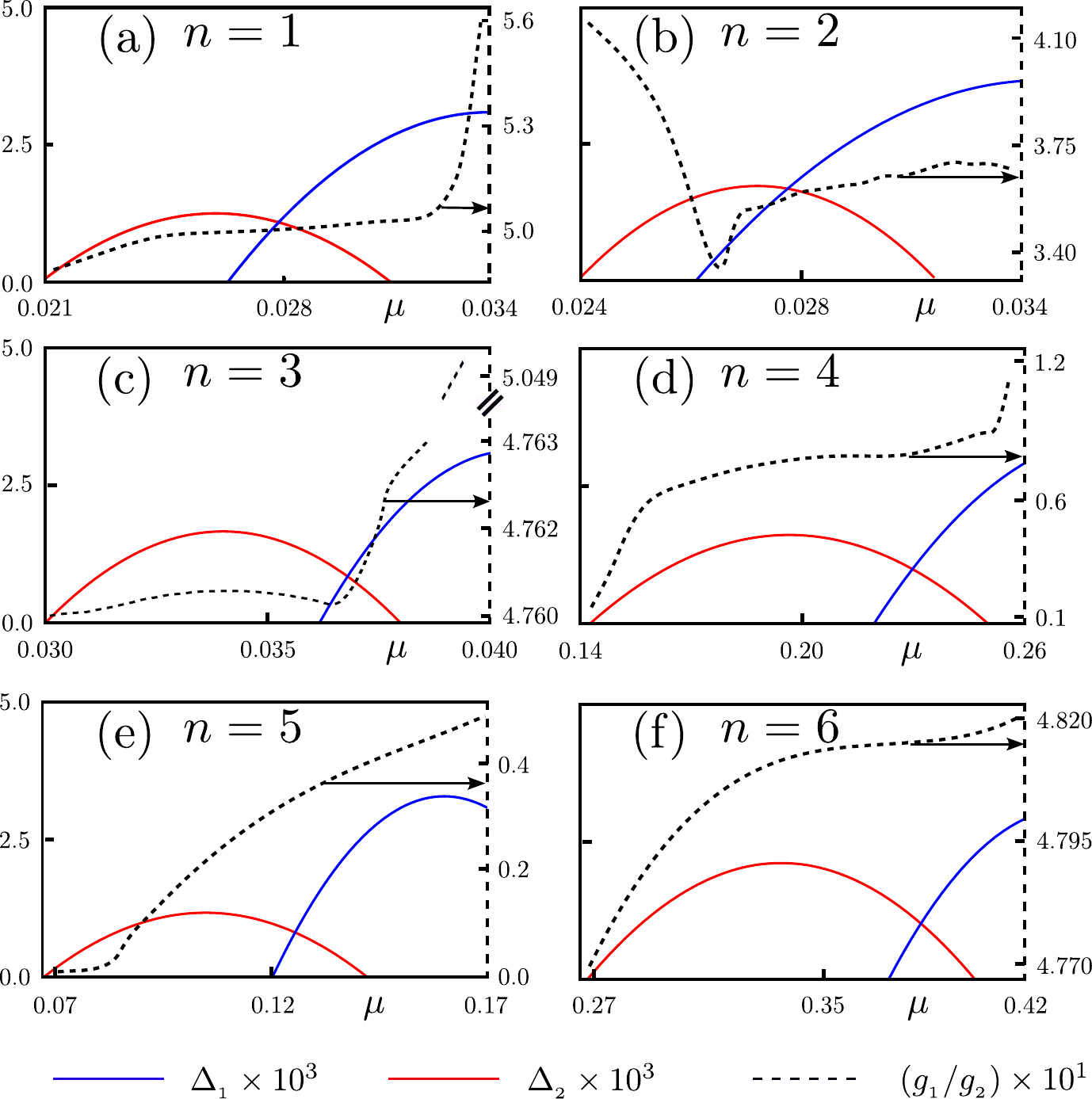}
    \caption{Solutions of the mean-field gap equation [obtained from Eq.~\eqref{eq:freeenergy}] at a temperature $T=10^{-5}$ displaying a competition between the anomalous Hall order with amplitude $\Delta_1$ (blue) and valley-coherent charge density wave with amplitude $\Delta_2$ (red) with numbers on left vertical axes as a function of the ratio of the corresponding coupling constants $g_{_1}$ and $g_{_2}$ (black dashed curve with numbers on right vertical dashed axes), respectively, over a range of chemical potential $\mu$ in the quarter metal of chirally-stacked $n$ layer graphene (see legends)~\cite{SM}. For smallest (largest) value of $\mu$ the system yields annular (simply connected) Fermi ring when $2 \leq n \leq 6$. For $n=1$ the Fermi ring is always simply connected. Appearance of anomalous Hall (valley-coherent charge density wave) order for large (small) chemical potential is chosen to qualitatively mimic the experimental observation in rhombohedral trilayer ($n=3$) and hexalayer ($n=6$) graphene. Here, $\Delta_{1}$, $\Delta_{2}$, and $T$ are measured in units of $t_0$ (Fig.~\ref{fig:lattice}).         
    }~\label{fig:meanfield}
\end{figure}

We, however, notice that besides masses there exists another class of orders that only \emph{partially} anticommute (or commute) with the free-fermion Hamiltonian. Although such orders typically tend to be energetically inferior to mass orders (following the second Clifford algebraic criterion), strong electronic interactions in the suitable channels can, in principle, cause their dynamic nucleation. Motivated by recent experimental observations in the quarter metal of rhombohedral trilayer and hexalayer graphene, reporting a second phase devoid of any $\sigma_{xy}$ and hysteresis in $R_{xy}$~\cite{ref:RHLG:1, ref:RTLG:5, ref:RTLG:6, ref:BBLG:1, ref:BBLG:2, ref:BBLG:3, ref:BBLG:4, ref:RTLG:2, ref:RTLG:3, ref:RTLG:4, ref:TTLG:1, ref:TTLG:2, ref:TTLG:3, ref:TTLG:4, ref:RPLG:1, ref:RPLG:2, ref:RPLG:3, ref:RHLG:5, ref:RHLG:6, ref:RHLG:7, ref:RHLG:8}, we re-search for its additional candidates. But, we stick to the guiding principle of the first Clifford algebraic principle, demanding that the corresponding matrix operators must \emph{commute} with ${\boldsymbol \sigma} \otimes \Gamma_{03}$ and $\sigma_0 \otimes \Gamma_{03}$. We immediately identify \emph{four} candidate matrix operators $\Gamma_{j0}$ and $\Gamma_{j3}$ with $j=1,2$. Such an order breaks the translational symmetry, generated by $\Gamma_{30}$ as $\{ \Gamma_{j0/j3},\Gamma_{30} \}=0$ for $j=1,2$ and mixes two valleys (appearing with the off-diagonal Pauli matrices $\tau_{1}$ and $\tau_{2}$ in the valley space). Hence, the resulting state is formed via a \emph{coherent} superposition of the fermionic states localized near two valleys. Now depending on $\eta_0$ or $\eta_3$, we find either in-phase or out-of-phase fermionic density modulations between the layers/sublattices with a characteristic periodicity of $2 {\bf K}$, as shown in Fig.~\ref{fig:orders}(a). Such an order causes an enlargement of the unit cell, yielding a CDW~\cite{SM, ref:RTLGnew:2, ref:MLGnew:1}. Hence, this phase is named VC-CDW.

Furthermore, the VC-CDW displays an intriguing property depending on the number of chirally-stacked honeycomb layers ($n$). Notice that for $n=2,4,\cdots$ ($n=3,5,\cdots$) the low-energy sites form an emergent honeycomb (prism-like) lattice (Fig.~\ref{fig:lattice}) for which the generator of the $C_3$ symmetry is $\Gamma_{33}$ ($\Gamma_{00}$). Hence, the VC-CDW order additionally breaks the three-fold rotational symmetry when $n=1,2,4, \cdots$ as $\{ \Gamma_{j0/j3}, \Gamma_{33} \}=0$ for $j=1,2$ thereby representing a stripe order therein, as reported for rhombohedral hexalayer graphene~\cite{ref:RHLG:1}. By contrast, the same phase preserves the $C_3$ rotational symmetry (as its generator is \emph{trivial} $\Gamma_{00}$) when $n=3,5,\cdots$ as observed in rhombohedral trilayer graphene~\cite{ref:RTLG:5, ref:RTLG:6}.

It should be noted that the first Clifford algebraic criterion allows another order that can lift the residual valley degeneracy of the electronic bands, namely the valley or chiral chemical potential, represented by the Hermitian matrix $\Gamma_{30}$, which, however, commutes with all the matrices appearing in the momentum-dependent kinetic terms $\Gamma_{kj}$ with $k=0,3$ and $j=1,2$. Consequently, such an order is energetically inferior to both AHO and VC-CDW, according to the second Clifford algebraic criterion, possibly for which it has not been observed in any experiment, otherwise fostering magneto-electric effects.

\subsection{Competition and coexistence}

Finally, we address a competition between these two ordered phases in the quarter metal phase within a mean-field approximation. Then, the free-energy density reads
\begin{eqnarray}~\label{eq:freeenergy}
    F= \sum_{j=1,2} \frac{\Delta_j^2}{2g_{_j}} - k_B T \sum_{\tau,\rho=\pm}\int_{\vec{k}} \ln{\left[\cosh{\left(\frac{|E^\rho_\tau-\mu|}{2 k_BT}\right)}\right]},\:\:\:
\end{eqnarray}
where $k_B$ is the Boltzmann constant, which we set to be unity, $\int_{\vec{k}} \equiv \int d^2{\vec k}/(2\pi)^2$, $\Delta_1$ ($\Delta_2$) is the amplitude of the AHO (VC-CDW) with the corresponding coupling constant $g_{_1}$ ($g_{_2}$), and $E^\rho_\tau$ are the eigenvalues of the effective single-particle Hamiltonian $H_n(\vec{k}) + \Delta_1 \Gamma_{33} + \Gamma_{10} \Delta_2$ in the presence of these two orders in the spin-polarized half metal phase. The coupling constants $g_{_1}$ and $g_{_2}$ are treated as phenomenological inputs whose strengths depend on the chemical potential $\mu$ (serving as an infrared cutoff) in the spirit of the renormalization group (RG), where these quantities are running variables. Minimizing $F$ with respect to $\Delta_1$ and $\Delta_2$, we arrive at coupled gap equations, which are solved numerically to mimic the experimentally observed phase diagrams in rhombohedral trilayer and hexalayer graphene, suggesting appearance of the VC-CDW (AHO) at lower (higher) doping in quarter metal~\cite{SM}. As the matrix operators for these two orders mutually \emph{anticommute}, their amplitudes enter the quasiparticle energy spectrum as sum of the squares. Hence, a coexistence between these two orders is energetically favored and we generically observe a regime of their coexistence that separates two pure phases (according to the third Clifford algebraic criterion). These features are shown in Fig.~\ref{fig:meanfield}, which we construct by taking parameter values from various experiments~\cite{SM}.

Even though we choose the phenomenological coupling constants $g_{_1}$ and $g_{_2}$ such that the AHO (VC-CDW) order appears in the higher (lower) doping regime, which qualitatively agrees with the experimental observations~\cite{ref:RHLG:1, ref:RTLG:5, ref:RTLG:6}, this outcome is also possibly generic, and can be \emph{qualitatively} justified by combining the second Clifford algebraic criterion and the spirit of RG. As the mass orders, such as the AHO, are energetically more favored over the others, such as VC-CDW, the strength of the bare microscopic interaction in the former sector is typically strongest among all possible symmetry breaking channels. Therefore, in order to reach the requisite strength to manifest a symmetry breaking, the \emph{scale-dependent} or \emph{running} coupling constants in the non-mass channel require more RG time compared to those in the mass channels. As the chemical potential sets an infrared cutoff for the RG flow of running interaction coupling constants, $\mu$ thus needs to be smaller to favor the condensation of non-mass orders, such as VC-CDW, while mass orders, such as AHO can be realized at a relatively high $\mu$, which are consistent with our phenomenological inputs in the mean-field analysis and experimental observations.

Since no ferromagnetic order has been observed in the half metal, here we do not study its competition with the antiferromagnet. However, at least algebraically the possibility of realizing a ferromagnetic order in a spin-polarized half metal cannot be ruled out. Otherwise, a competition between these two orders leads to similar outcomes we find from the confluence of AHO and VC-CDW. Namely, besides pure ferromagnetic and antiferromagnetic orders, which are expected to appear in the lower and higher doping regimes in the half metallic phases (following the above argument combining the second Clifford algebraic criterion and the spirit of RG), respectively, we expect a regime of coexistence between them. Fascinatingly, in the regime of coexistence, the ferromagnetic order develops magnetization in a specific direction, called the easy-axis, while the Ne\'el vector gets confined to the easy-plane, perpendicular to the direction of the magnetization; also named \emph{canted antiferromagnet}, such that the respective matrix operators, namely $\sigma_3 \otimes \Gamma_{00}$ and $\vec{\sigma}_\perp \otimes \Gamma_{03}$ mutually anticommute (third Clifford algebraic criterion), where $\vec{\sigma}_\perp=(\sigma_1, \sigma_2)$.

\subsection{Valley-polarized half metal}

Even though the first Clifford algebraic criterion selects unique sets of orderings for the spin and valley degeneracy lifting, it is a priori unclear which degeneracy gets lifted first at relatively higher doping yielding a half metal, followed by the appearance of a fully-polarized quarter metal at even lower doping, without any microscopic knowledge. A reasonable assumption is that the on-site Hubbard repulsion ($U$) is possibly the dominant component among all the finite range electronic interactions and the intra-layer next-nearest-neighbor repulsion ($V_2$) is the next dominant channel, which immediately leads to the formation of a spin-polarized half metal with antiferromagnetic order and fully-polarized quarter metal with additional AHO (or sometimes VC-CDW), as typically observed in $n$ layer chirally-stacked multilayer graphene heterostructures, subject to $D$ field, causing layer (or sublattice) polarization~\cite{ref:RHLG:1, ref:RTLG:5, ref:RTLG:6, ref:BBLG:1, ref:BBLG:2, ref:BBLG:3, ref:BBLG:4, ref:RTLG:2, ref:RTLG:3, ref:RTLG:4, ref:TTLG:1, ref:TTLG:2, ref:TTLG:3, ref:TTLG:4, ref:RPLG:1, ref:RPLG:2, ref:RPLG:3, ref:RHLG:5, ref:RHLG:6, ref:RHLG:7, ref:RHLG:8}
.

However, the possibility of a \emph{valley-polarized half metal} cannot be ruled out, at least algebraically. In fact such a phase has been observed in hole-doped pentalayer rhombohedral graphene ($n=5$) for a weaker $D$ field, displaying hysteresis in $R_{xy}$~\cite{ref:RPLG:3}, strongly suggesting an AHO therein, compatible with our first (primarily) and second Clifford algebraic criteria. However, such a phase is realized for very low carrier density thereby possibly forbidding the realization of a fully polarized quarter metal, realized by subsequent nucleation of antiferromagnet or ferromagnet. Rather in the low electric field regime, the system fosters a correlated insulator at and near half-filling, which according to our Clifford algebraic criteria should be spin and valley polarized that can be tested in future experiments measuring magnetic hysteresis. Since in pentalayer rhombohedral graphene the opposite layers, fostering low-energy quasiparticles, are sufficiently far ($\sim 1.5$ nm), hysteresis can possibly be measured on the top and bottom layers separately. With the layer antiferromagnet (ferromagnet) in such a correlated insulator, the magnetic hysteresis on the opposite layers should be of opposite (same) sign, which can be a good test of our proposed theoretical picture, based on Clifford algebraic arguments. It is also conceivable that such a correlated insulator is devoid of any spontaneous symmetry breaking (such as Mott insulators). We hope that the present work initiates such future experiments.

\section{Discussions and future directions}~\label{sec:discussions}

To conclude, from a Clifford algebraic universal protocol (the first one) for the systematic degeneracy lifting of electronic bands, we identified two candidate orders for nondegenerate quarter metals in chirally-stacked $n$ layer graphene, namely VC-CDW and AHO. The VC-CDW (AHO) mixes two valleys (causes valley polarization) and thus breaks (preserves) the transitional symmetry. The hallmark anomalous $\sigma_{xy}$ and hysteresis of $R_{xy}$ in AHO have been observed in the quarter metal of all the systems with $2 \leq n \leq 6$~\cite{ref:RHLG:1, ref:RTLG:5, ref:RTLG:6, ref:BBLG:1, ref:BBLG:2, ref:BBLG:3, ref:BBLG:4, ref:RTLG:2, ref:RTLG:3, ref:RTLG:4, ref:TTLG:1, ref:TTLG:2, ref:TTLG:3, ref:TTLG:4, ref:RPLG:1, ref:RPLG:2, ref:RPLG:3, ref:RHLG:5, ref:RHLG:6, ref:RHLG:7, ref:RHLG:8}. However, thus far the VC-CDW ordering has been observed in the quarter metal of rhombohedral trilayer and hexalayer graphene, where the $C_3$ symmetry is respectively preserved~\cite{ref:RTLG:5, ref:RTLG:6} and broken~\cite{ref:RHLG:1}. Such an outcome is consistent with the second Clifford algebraic criterion. Thus, the VC-CDW order corresponds to a stripe phase only in the latter system, in agreement with our theoretical findings. As disorder (also breaking the translational symmetry) couple with VC-CDW as a \emph{random field}, their inevitable presence in real materials can be detrimental for the nucleation of such an order~\cite{ImryMa}. Nonetheless, we are optimistic that with increasing sample quality, the VC-CDW order can be observed in other chirally-stacked graphene heterostructures. Although an exact microscopic interaction for the VC-CDW is currently unknown, which is a typical situation in strongly correlated materials, future works should be be devoted to search for it.

Throughout this work, we have neglected the long-range tail of the Coulomb interaction as due to the constant (for $n=2$) or diverging (for $n \geq 3$) density of states, it gets screened. Even in monolayer graphene, supporting massless Dirac fermions for which it is completely unscreened and goes as $1/|\vec{k}|$, the fully unscreened Coulomb interaction does not support any broken symmetry phases, as shown from extensive
diagramatic Monte Carlo calculations~\cite{DMC:Prokofiev}. Therefore, in multilayer graphene with non-linear band dispersion when Thomas-Fermi screening converts long-range Coulomb interaction into a short-range one, it does not play any role in symmetry breaking. Rather the screened Coulomb interactions only changes the bare values of various short-range components of the electronic interaction, which are responsible for various ordering. Such an outcome does not change even when trigonal warping is accounted for, yielding Dirac points at the lowest energy scale for which the long-range Coulomb interaction is unscreened.

Even though no ordering has been observed in pristine monolayer graphene so far, recent experiments have revealed relativistic Mott transitions into a mass ordered state in designer graphene~\cite{ref:MLG:4, ref:MLG:5}. In such a system a similar cascade of degeneracy lifting can be observed with the following protocol. The analog of layer polarization can be engineered by creating a small density imbalance between two sublattices. On such an engineered band-insulating system, one should observe a cascade of degeneracy lifting when it is doped away from the CNP, similar to the situation in chirally-stacked graphene multilayer. However, due to the $E$-linear density of states, the on-site and next-nearest-neighbor repulsions need to be stronger, which can possibly be achieved in designer systems. Optical honeycomb lattices constitute yet another promising platform where one can observe the cascade of degeneracy lifting and formation of fractional `thermal' metals of neutral atoms, where Hubbard repulsion-driven antiferromagnetism has already been emulated~\cite{optgra:1}. Thus with an externally induced sublattice staggered potential, the realization of thermal half metal should be within the reach of existing experimental facilities. Even though tuning next-nearest-neighbor repulsion to a desired strength to lift the residual valley degeneracy can be challenging therein, the requisite AHO and VC-CDW can be engineered externally, as has been achieved with the former one~\cite{optgra:2}. Therefore, with engineered AHO or VC-CDW, a thermal quarter metal on Hubbard-dominated optical honeycomb lattices can be observed with a staggered sublattice potential.

Our proposed Clifford algebraic protocol for systematic band degeneracy lifting should be generically applicable in any multi-band system, among which twisted bilayer graphene is the most prominent one. However, due to sufficiently narrow band width close to the so-called magic angle~\cite{TBLG:0}, such a system usually becomes an insulator upon degeneracy lifting at various integer fillings, as observed in experiments~\cite{TBLG:1, TBLG:2, TBLG:3, TBLG:4}. A detailed analysis on such a system is left for a future investigation.

\acknowledgments

This work was supported by NSF CAREER Grant No.\ DMR-2238679 of B.R.\ (S.A.M.\ and B.R.) and Dr.\ Hyo Sang Lee Graduate Fellowship from Lehigh University (S.A.M.). We thank Vladimir Juri\v ci\' c, Christopher A.\ Leong, and Daniel J.\ Salib for critical input on the manuscript. We thank Jia Leo Li for useful correspondences. 

\section*{Data availabity}

The numerical codes and data that support the findings of this article are openly available in Ref.~\cite{codesdata}

\end{document}